\documentclass[11pt]{article}
\usepackage{moriond,epsfig}

\def\be{\begin{equation}}
\def\ee{\end{equation}}
\def\bea{\begin{eqnarray}}
\def\eea{\end{eqnarray}}

\begin{document}
\vspace*{4cm}
\title{Secular Evolution of Galaxies}

\author{ F. Combes }

\address{LERMA, Observatoire de Paris,\\
61 Av. de l'Observatoire, F-75014 Paris, France}

\maketitle\abstracts{
In current $\Lambda$CDM galaxy formation scenarios, at least three 
physical phenomena could contribute to the mass assembly: 
monolithic collapse, hierarchical mergers and more quiescent 
external gas accretion, with secular evolution. 
The three processes are described, and their successes and
problems are reviewed.  It is shown that monolithic 
collapse is likely to be quite restricted to sub-components
of galaxies, while the two main scenarii, hierarchical
merging and secular evolution, might have comparable
roles, depending on environment.  Evidences are reviewed
for the important role of gas accretion,
followed by secular evolution.  In particular the
existence of thin and cold disks, the occurence of
bars and spiral structure, the frequency of lopsided
instabilities, and the history of star formation, all
point towards large amounts of cold gas accretion.
Some examples of N-body simulations are reviewed in support
of secular evolution.
}

%--------------------------------------------------------------------------
\section{The three main processes/scenarios of galaxy formation} 

\subsection{Monolithic Collapse (MC)}

The scenario of monolithic collapse for the formation of the Galaxy
was suggested by Eggen, Lynden-Bell and Sandage (1962, ELS),  on the basis
of the observed correlation between [Fe/H] and orbital 
excentricity $e$  for old stars.  Since stars with low metallicity
had very low angular momentum L$_z$, they suggested that the old stars 
were formed out of gas falling towards the center in radial orbits, 
collapsing quickly from a halo to a thin rotating disk plane
enriched in heavy elements by star formation. In this picture,
the gravitational potential is varying slowly and the stellar parameters 
$e$ and L$_z$ can be considered as invariants.

In an opposite view,  Searle \& Zinn (1978) consider that the 
ELS results are simply the consequence of the selection bias on high proper motion stars.
They remark that halo globular clusters have a large range of metallicity, with
 [Fe/H]   uncorrelated with distance, and favor a formation through merging of small protogalaxies.
The gravitational potential is varying rapidly in this view, and there is also
substantial gas accreted later.

The monolithic collapse scenario for our Galaxy is
not viable anymore, or only for a restricted component,
the central bulge.  The stellar halo is smaller than the disk,
and could not collapse into the disk. The only galactic 
component to collapse into is the old bulge (Gilmore 1996).
Note however that, although the bulge stars are old, they have
high metallicity, as the disk, so this is not easily explained
by this scenario.

Rapid monolithic collapse is not likely to be the main
formation process of most normal galaxies, 
since huge ULIRGs at high redshift  are very rare. In any case,
departures from MC  are larger in low-mass galaxies 
(Ferreras \& Silk 2000).

\subsection{Hierarchical scenario (HS)}

 In this frame, a system like our own Galaxy is the result of the hierarchical
assembly of dark halo building blocks. Accretion of baryonic gas
occurs later, in the assembled structure, to form the bulge, and progressively the thin disk,
which forms last. The thick disk could be due to the past heating of the thin disk by
companions. 

This scenario is supported by the existence of stellar streams in the stellar halo
(Helmi 2002, Ibata et al 2002): they are tidal debris from the accretion
of small companions. The halo could be entirely built from
these minor mergers.

The observed large increase with redshift of the merger frequency
comes in support of the hierarchical scenario. This is particularly spectacular
in galaxy clusters, for massive galaxy formation (e.g. van Dokkum et al 1999).
In some massive ellipticals, two populations of globular clusters 
have been detected, hinting towards two major merging episodes,
triggering their formation (Zepf \& Ashman 1993).

\subsection{Secular Evolution, with slow and continuous external matter accretion (SE)}

In secular evolution, the bulge component is formed slowly from the disk
through the bar action, and the disk can be replenished through continuous
external gas accretion.  The Galaxy can be considered an open system,
with slow mass growth through time, from external accretion of gas
progressively transformed into stars, and the various components
interacting with each other.

According to a statistical analysis of 257 spiral galaxies, the radial color gradients
observed and the relation between disk and bulge colors favor the
formation of bulges through SE (Gadotti and dos Anjos 2001). SE is
also supported by the relation between bulge and disk masses and radii
(Courteau et al 1996).

In a large sample of early-type galaxies, 
moderate radial gradients of metallicity are observed,
which are independent of luminosity
(De Propris et al 2005). Ellipticals and bulges of lenticular
galaxies reveal similar behaviour.  These observations
are contrary to what is expected with MC (large gradients 
which should increase with mass and luminosity)
and contrary to what is predicted by HS (which smears out gradients).

Cosmological simulations based on a $\Lambda$CDM universe
may include all processes with different emphasis, according
to the treatment of baryon physics. In addition to hierarchical
merging of dark haloes, the hot gas is cooling in dark haloes
within the cooling radius, and contracts monolithically to form galaxies.
However, the gas infall might also occur in fragmented and dense
colder gas clouds (Maller \& Bullock 2004), or the gas accretion could
be continuous, and favor secular evolution.

Successive episodes of dissipational collapse, followed by
mergers, lead in numerical simulations to elliptical-like
objects, which are in better agreement with observations than
the results of stellar disks mergers only (Meza et al 2003).
In this scenario, the final merger episode forms hot gas, 
the absence of cooling prevents star formation, and the final
object is an elliptical, with no recent accretion, and 
consequently an old stellar population with no disk morphology.

%--------------------------------------------------------------------------
\section{Successes and difficulties of the different processes}

\subsection{Massive galaxies early in the universe}

Very massive elliptical systems (10$^{11} M_\odot$), possessing a large
fraction of old stars, have been observed in deep optical and near-infrared surveys,
at redshifts between 1 and 2 (Daddi et al 2004, Cimatti et al 2004). The presence of these rather
evolved systems, when the universe was only 25\% of its age was a surprise,
since it is expected that massive ellipticals are the end result of the hierarchical
merging process.

However, these massive objects are highly clustered and appear to correspond 
to the early formation of groups in rich environments. It is  predicted that
evolution proceeds much more rapidly in those environments with
respect to the field, and the presence of some fraction of massive
objects is thus expected.

Such massive objects can be traced by their star formation, 
either by optical techniques, when their light is not highly
obscured, as in Lyman Break Galaxies (LBG), or through their dust
emission in the far-infrared, redshifted in the submillimeter range.
The latter are then called SMGs (SubMillimeter
Galaxies), and are found preferentially between z=1 and 4.
The physical nature of these objects can be determined more precisely
when CO emission is detected.  Given their gas content, and star formation
rate (larger than 700 M$\odot$/yr), their life-time is between 40 Myr
and 200 Myr (Greve et al 2005). The correlation between the Far-Infrared
and CO emission suggests that they are the equivalent of local ULIRGs, with
an even more efficient star formation rate.
These starbursts are associated to galaxy mergers, which could be more violent
than locally, since bulges are less prominent at high redshifts.

\begin{figure}[ht]
\centerline{\includegraphics[width=13cm]{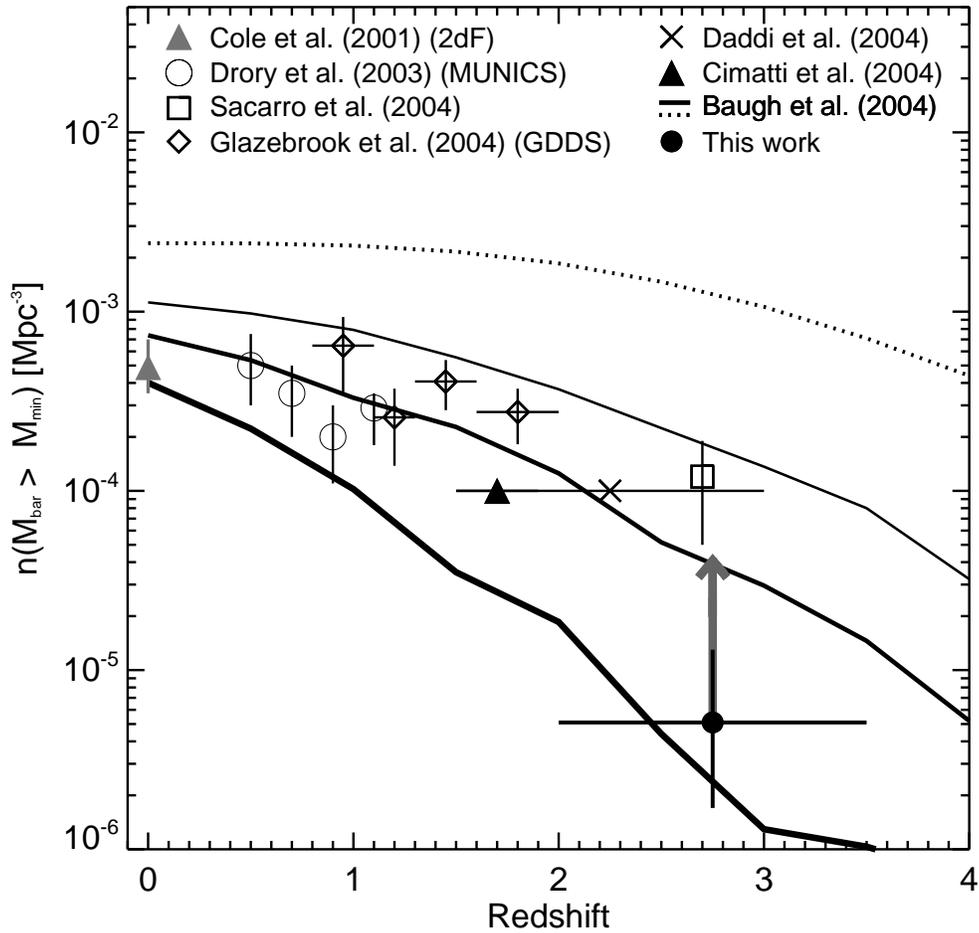}}
\caption{The co-moving number density of galaxies with visible
masses $\sim 6\times 10^{10}\, M_\odot$
as derived from CO observations of SMGs. The predictions from the
recent semi-analytic model from Baugh et al. (2005) 
of the abundances of galaxies with baryonic masses $\ge 5\times$, $7\times$,
and $10\times 10^{10}\, M_\odot$ as a function of redshift are shown
respectively as thin, medium and thick solid lines.
The dotted curve represents the total baryonic matter content available
in $\ge 10^{11}\, M_\odot$ dark matter halos, and is obtained
by scaling the abundance of dark halos with 
$\Omega_b/\Omega_{DM} = 0.13$. 
The observational points are compatible with the model predictions,
provided that $\sim 10$ per cent of all baryons within dark matter halos are
rapidly assembled into galaxies.}
\label{greve05}
\end{figure}

The number of SMGs detected can be used to estimate the 
density of massive galaxies at these redshifts 
(Greve et al 2005). The estimate is a lower limit, since
the life-time of the starburst phase is not precisely
known (and the correction is made with the conservative value of
200 Myr). Within the error bars, the derived number density
of SMGs is compatible with the predictions of the 
$\Lambda$CDM simulations and semi-analytical calculations, 
if 10\% of baryons are rapidly converted into galaxies
(cf Figure \ref{greve05}).

This agreement with observations is obtained in the new model
from Baugh et al (2005)
through a few changes in the assumptions of the semi-analytical 
simulations, in particular a much larger influence of bursts
in the star formation history, a longer time-scale for 
star formation in disks, and most important, the assumption
of a top-heavy IMF in starbursts at high redshift. 
The consequences of the new model are that the star formation is dominated
by bursts at redshifts larger than 4, and later by quiescent mode
(while it was dominated by the quiescent mode at all redshifts before).
Integrated over the Hubble time, bursts are now responsible
for 30\% of all star formation (Baugh et al 2005).
 Massive galaxies today have only a few percent of their stars
formed in bursts at z $\sim$ 2. The galaxy formation is still
quite far from the monolithic collapse view, since the bulk of the
stars in massive ellipticals is formed in disks, and then
re-arranged in spheroids during mergers.
Note that the estimated baryonic mass of SMGs has now been
re-scaled to 0.6 M$_*$ in average, while it was 4 times larger
before (Greve et al 2005).

\subsection{Star formation history versus mass}

The star formation history (SFH) has been derived from the direct observations of
young stars in distant galaxies at different redshifts, which reveal a peak in the 
star formation rate about 8 Gyr ago, and then a decline by a factor of $\sim$ 10.
Alternatively, the study of the fossil record of stellar populations, in the local
large sample of galaxies in the SDSS has also led to a similar star formation
history, with a peak occuring slightly later, about 5 Gyr ago
(Heavens et al 2004, Jimenez et al 2004). This study also shows that the SFH is
different according to the mass. Stars in
massive galaxies appear to have formed at early times, and these galaxies are
not actively forming stars now, while dwarf galaxies appear to experience
starburts.
Only intermediate masses have in average maintained
their star formation rate over a Hubble time.

It has been known for a long time that
galaxies in the middle of the Hubble sequence had about
constant SFR across the Hubble time (Kennicutt 1983, Kennicutt et al 1994).
Even taking into account the stellar mass loss, an isolated
galaxy should have an exponentially decreasing  star formation history.
 To account for observations, a source of gas should be found
to replenish the star formation fuel in such galaxies.
 This cannot come from major mergers, which destroy disks, and 
lead to early-type or elliptical galaxies.  On the other hand,
small companions are not sufficient, for instance systems falling
now on the Milky Way (Sag dw, Canis major, etc..) are of the order of
1/400th of the mass of the Galaxy (Ibata et al 2001, 2003).
Accretion of gas from the cosmic filaments should provide
the required fuel, which corresponds to a slow and continuous
secular evolution.

\subsection{Quiescent star formation versus bursts}

Below a redshift of z =0.7, which corresponds to about half of the
Hubble time, the star formation rate of the universe
is dominated by a rather quiescent mode in normal galaxies. 
But at higher redshift, starbursts begin to dominate, first
moderate bursts, giving rise to LIRGs (Luminous Infra-Red Galaxies,
with L $> 10^{11} L_\odot$), and after z=4, ULIRGs (Ultra-Luminous
Infra-Red galaxies, with L $> 10^{12} L_\odot$).
Given the shape of the star formation history, most of the stars today
have been born in LIRGs (63\%), the rest being in ULIRGs (21\%)
or normal galaxies (16\%); this is consistent with the cosmic
infrared backgound (CIRB), which is dominated at least within a
proportion of two thirds by LIRGs (Elbaz \& Moy, 2004).

The fact that
LIRGs are more numerous at z $>$ 0.4 (15\%) than today ($\sim$ 1\%)
has been interpreted in terms of recent formation (in the 
second half of the universe) of the bulk
of stars in intermediate-mass galaxies (Hammer et al 2005).
At this epoch, LIRGs can account for 38\% of all star formation.
The high frequency of LIRGs  implies episodic star formation 
bursts, maybe triggered by galaxy interactions.
The starburst phase is of the order of 100 Myr, while an interaction
is of the order of 10$^9$ yrs. This means that all galaxies should be
interacting at z$>$ 0.4, which is not what is observed.

The frequency of interactions must be lower,
but another process could compensate: external gas accretion
from the cosmic filaments.  Secular evolution could then be
responsible for these episodes of star formation. In any scenario,
gas must be provided to replenish spiral disks, whatever is
the final dynamical trigger, either a passing companion or
internal evolution.
In both cases, the amount of fuel is the same, the gas from the disk 
of one (or two) galaxy, and a bar drives the gas towards the center. 
The gravity torques from the bar can fuel the central disk
and star formation intermittently, through self-regulated
mechanisms (see below).

\subsection{Late bulge formation}

Contrary to massive spheroids, consisting of old and red stellar populations,
many bulges of spiral galaxies appear to have formed later than their disk.
They host young populations, that must have been formed from recent gas inflow
followed by star formation, due either to secular evolution or to
galaxy interactions. Using the criterion that the color inside the half-mass
radius is bluer than the outer parts, Kannappan et al (2004) found that
10\% of galaxies in the NFGS (Nearby Field Galaxy Survey) are experiencing
bulge growth at the present time.

\begin{figure}[ht]
\centerline{ \begin{tabular}{cc}
\includegraphics[angle=0,width=7cm]{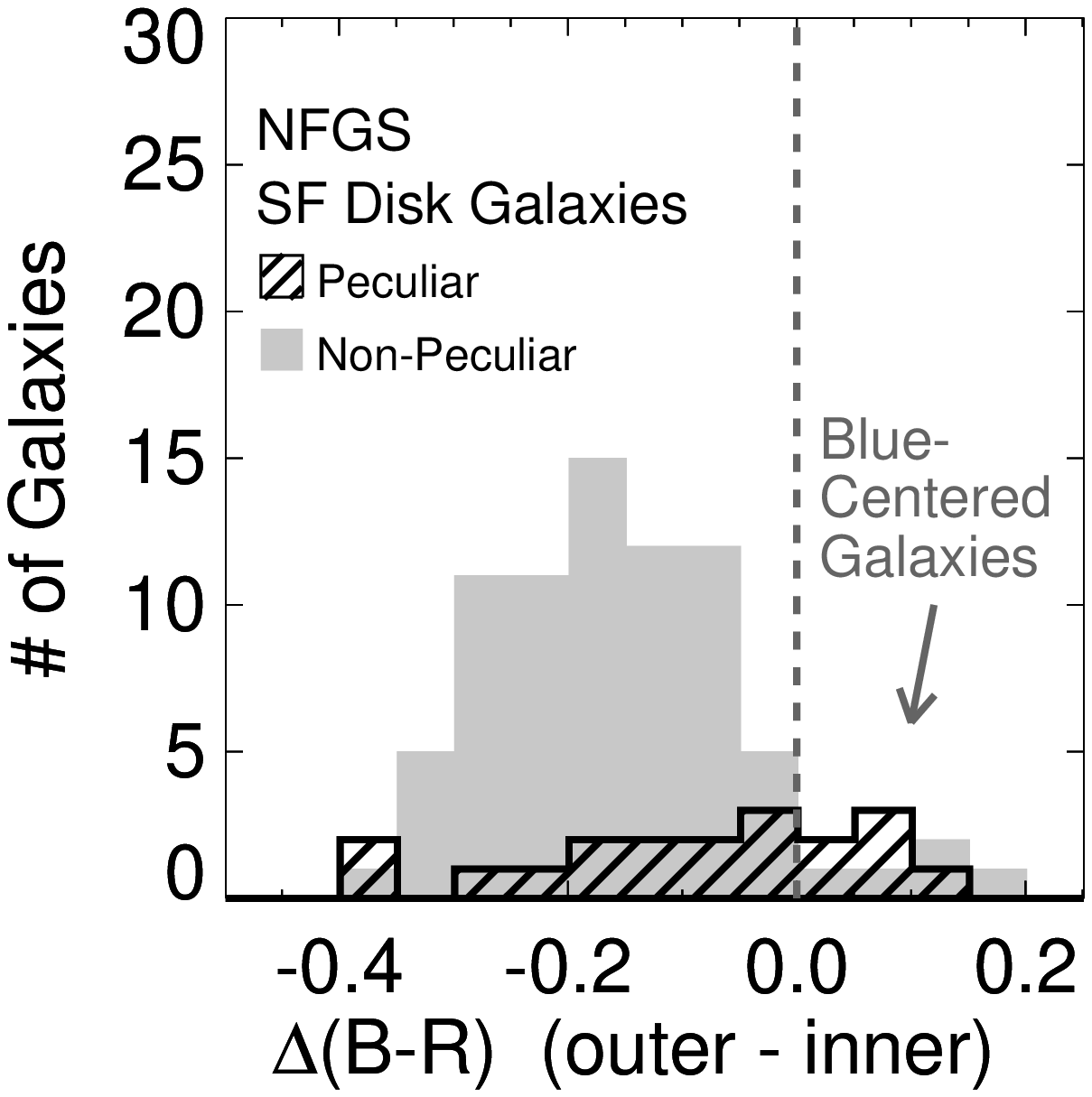}&
\includegraphics[angle=0,width=7cm]{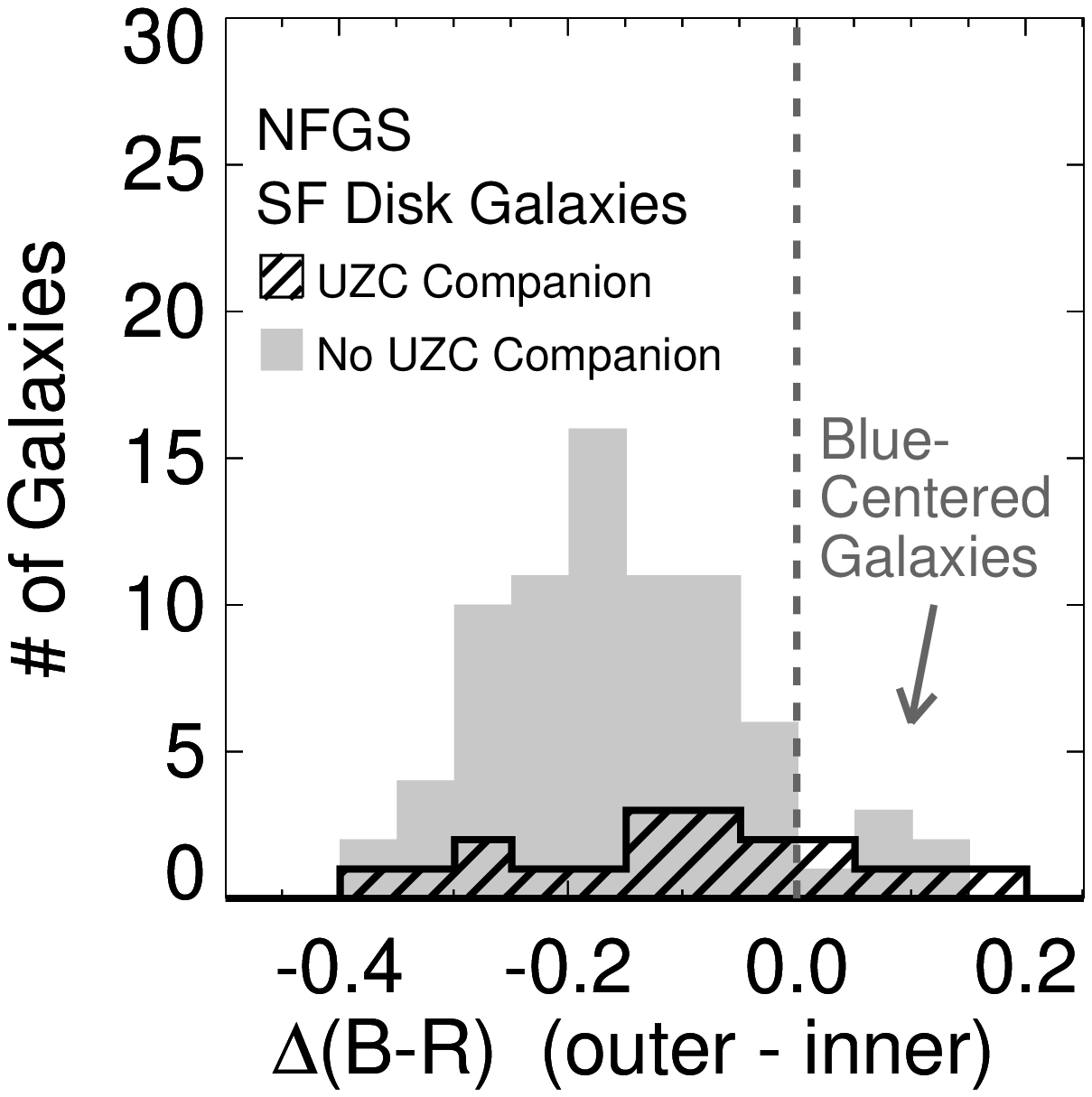}\\
\end{tabular}}
\caption{Distribution of the
color difference $\Delta$(B-R) for star-forming disk galaxies {\bf (a)} with and
without morphological peculiarities, and {\bf (b)} with and without close
companions in the Updated Zwicky Catalog
(from Kannappan et al 2004)}
\label{kann}
\end{figure}

 To distinguish the actual source of gas inflow in those galaxies,
a correlation was searched between the blue-center disk characteristic
and the presence of companions (see Figure \ref{kann}). 
Although there is not a large correlation with the presence of
visible companions, there is a strong one with
perturbed morphology. The authors then conclude that external
drivers, minor mergers or interactions, are more likely the origin 
of bulge growth today, than internal secular evolution. In fact,
as will be developped later, secular evolution can also be driven
by external gas accretion, which if asymmetrical, can lead to
perturbed morphology as well.

\subsection{Respective role of MC, HS and SE}

In summary, a certain  amount of all three processes are
certainly present in galaxy formation, and it is quite
difficult to distinguish the relative role of each.
For instance, although MC is likely more effective at
high redshifts and in high-density environments, it is possible
that old stars have been accumulating in massive systems at 
various redshifts.

Galaxy interactions and mergers increase with $(1+z)^m$, with $m \sim$ 4.
It has been thus suggested that
hierarchical merging dominate in the past and secular evolution will in the future 
(Kormendy \& Kennicutt 2004).
But the availability of gas accretion also decreases with time, and it is quite
possible that today both processes are occuring with comparable importance,
depending strongly on environment. Both the effects of 
interactions and gas accretion are quenched in
clusters.  Even inside a given group, the past evolution appears to depend on 
morphological type: the
Milky Way reveals more signs of secular evolution (pseudo-bulge, 
old globular clusters..), while Andromeda has experienced a major merger
more recently.

\begin{figure}[ht]
\centerline{\includegraphics[width=16cm]{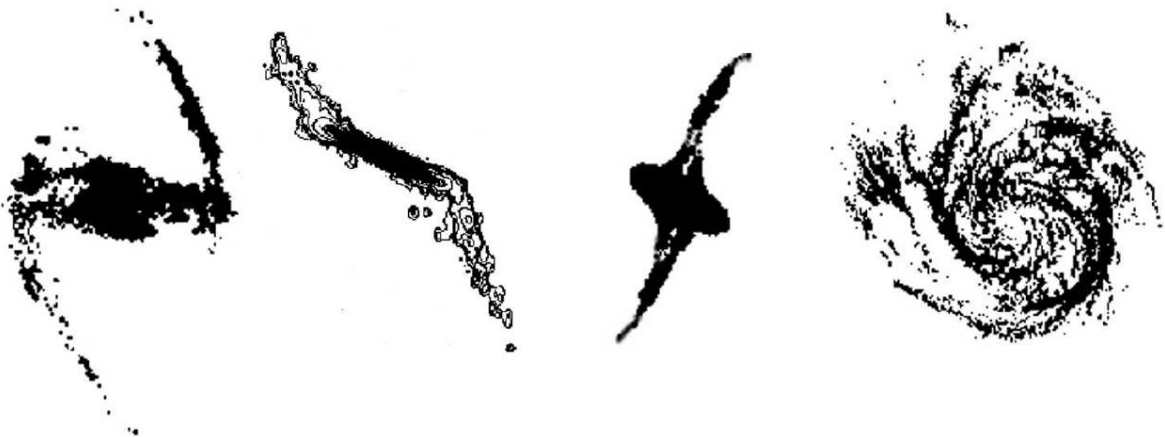}}
\caption{Dynamical processes able to constrain the rate of
external gas accretion, illustrated by prototypical examples,
from left to right: bars in galaxies (NGC 1365), 
warps (NGC 4013), polar rings (NGC 4650A), and lopsidedness (M101).}
\label{prgwb1}
\end{figure}

The perturbed morphology of galaxies should not
always be interpreted in terms of galaxy interactions, 
it could also be the result of external gas
accretion, which is most often asymmetric:
lopsided systems, warps, polar accretion...
(cf Figure \ref{prgwb1}).
External gas accretion, followed by secular evolution from the bar,
can also provide starbursts. The gravity torques of the bar 
either prevent or favor the gas infall towards the center,
as will be described now. The self-regulated bar action
results in intermittent periods of activity, 
starbursts ad well as nuclear activity.

%--------------------------------------------------------------------------
\section{Bars and secular evolution}

Dynamical instabilities in spiral disks 
are responsible for its evolution. Detailed numerical
simulations since several years have unveiled a
self-regulated cycle  in this secular evolution.
The cycle begins by a first bar instability
in a cold spiral disk with stars and gas.
When the bar is strong enough, it produces gravity torques
on the gas component, due to the phase shift of the gas
response (the gas being dissipative).
These torques drive gas inflow, from corotation to the center
(or ILRs, when they exist). The gas is losing its
angular momentum to the benefit of the bar,
which is then weakening.
The destruction of the bar is due essentially to the gas inflow
itself, and also to a lesser degree to the presence of a central
mass concentration (CMC), built in the process.
Only if enough external gas accretion 
occurs to replenish the disk mass and trigger
another bar instability, can the cycle loop again.

There have been debates about this cycle, in particular
about the efficiency of bar destruction, about their
ability to reform, and about the required central
mass (CMC) to weaken the bar.

\subsection{Role of gas in bar destruction}

Self-consistent simulations of spiral galaxies
with gas have shown that only the infall of 1-2\%
of mass in gas is sufficient to destroy the stellar bar
or to transform it into a lens
(Friedli 1994, Berentzen et al 1998, 
Bournaud \& Combes 2002, 2005). But the mechanism of
destruction was attributed to the formation of a central mass 
concentration (CMC), and simulations with the 
growth of an artificial CMC in the center of a few percent is not 
enough to destroy the bar (Shen \& Sellwood 2004).
Now, several simulations with and without gas or CMC
have clearly shown the role of gas in bar
destruction: 
gas is driven in by the bar gravity torques, and its
angular momentum is taken up by the bar wave.
In other words, the reciprocal torques from the gas
provides angular momentum to the bar, weakening it,
since the bar wave has negative angular momentum inside corotation
(Bournaud \& Combes 2005).
Since the formed CMC is not enough to destroy the bar by itself, it is
then more easy to reform a bar, when the disk has become unstable
again, through external gas accretion.

\begin{figure}[ht]
\centerline{ 
\includegraphics[angle=0,width=14cm]{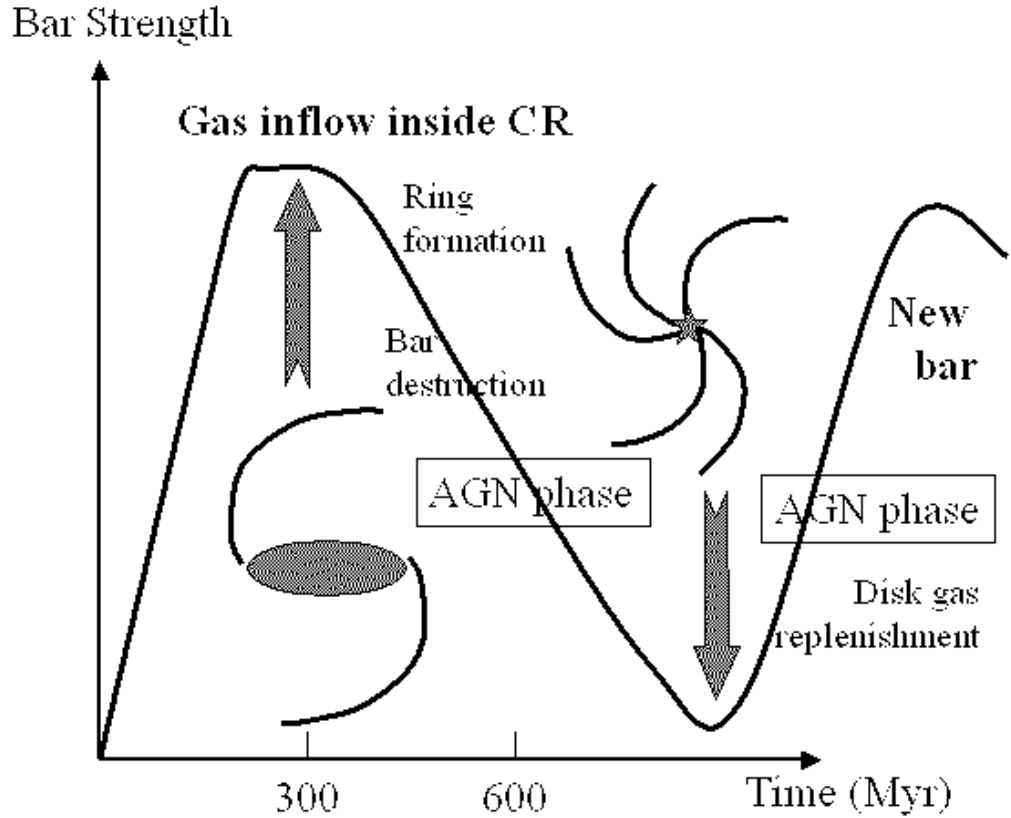}}
\caption{Schematic scenario of two bar episodes, and the
corresponding phases for the AGN fueling. When the bar grows,
it progressively drives gas infall inside corotation, while the 
accreted external gas is stalled at the OLR in the outer parts.
The gas then piles up at ILR (nuclear ring), but does not fall
into the center. It has to wait for the destruction or weakening of the bar,
at which stage viscous torques can fuel the AGN, and the external
gas can replenish the disk, to prepare another bar instability.}
\label{schema-bar}
\end{figure}

In this cycle, it is interesting to note that the galaxy
accretes gas towards its center by intermittence.
Indeed, while the bar is strong in the disk,
the gas from corotation to OLR (Outer Lindblad Resonance) 
is driven outwards by the positive gravity torques from the bar.
Gas is then stalled at the OLR in the 
border of the disk, prevented to enter.
The external gas remains there, while the gas inside
corotation is driven inwards, 
until the bar weakens. After the
bar destruction, the external gas can  enter
 and replenish the disk, to make it 
unstable again to bar formation
(see Figure \ref{schema-bar}).
This scenario explains why nuclear activity
is not well correlated with the presence of bars in galaxies,
even if the gas driven by bars fuels the AGN
(e.g. Garcia-Burillo et al 2005). In this frame,
AGN activity can be triggered in the weakening phases 
of the bar.

\subsection{Bar frequency}

Recent near-infrared surveys have allowed a statistical estimation
of bar strength in local galaxies
 (Block et al 2002, Laurikainen et al 2004). The observed
bar frequency is much larger than what is expected for isolated
galaxies, where bars should have been destroyed.
These statistics can be used to 
 quantify  the gas accretion rate, required to
reform bars in the right proportions.
  Note that the bar frequency has been observed
comparable at high redshift (Jogee et al 2004), and therefore
gas accretion must have played a role all along the Hubble time.

The gas is required to replenish the disk, and 
maintain it cold and unstable.  Not more than 10\% of the accretion
can be provided by dwarf companions, since 
interactions between galaxies heat the disk (Toth \& Ostriker 1992),
and massive interactions develop the spheroids.
The fitting of bar frequency between simulations and 
observations implies that a galaxy doubles its mass
through gas accretion in about 10 Gyr (Bournaud \& Combes 2002, 
Block et al 2002). The source of
 continuous cold gas accretion can come
from the cosmic filamentary structure
 in the near environment of galaxies.
Cosmological accretion in cosmological simulations 
confirm this gas accretion rate
(e.g. Semelin \& Combes 2005), and therefore bar reformation.

%--------------------------------------------------------------------------
\section{Warps and polar rings}

The presence of warps in almost all galaxies (conspicuous in HI-21cm)
can only be explained through misaligned external gas accretion 
(e.g. Binney 1992). The amount required can lead to the
reorientation of the angular momentum of the galaxy
in 7-10 Gyr (Jiang \& Binney 1999). This would correspond to the same
amount of gas accretion that can explain bar frequency.

Polar Ring Galaxies (PRG) are composed of an early-type host
surrounded by a perpendicular ring of young stars and gas,
 akin to late-type galaxy disks.
Stars in the polar ring have formed after
the interaction/accretion event, from the gas settled
afterwards in the polar plane.
The frequency of PRGs deduced from observations is
about 5\% (Whitmore et al 1990).

The formation of polar rings can be explained
either by gas accretion (Schweizer et al 1983,
Reshetnikov et al 1997), or by a galaxy merger (Bekki 1997, 1998).
External gas can be accreted either
from a passing by companion, 
or from the cosmic web filaments. Numerical simulations
reveal that it is about 5 times more
 probable to form a PRG by gas accretion
(Bournaud \& Combes 2003).

%--------------------------------------------------------------------------
\section{Lopsided galaxies}

The frequency of asymmetries in galaxies can
also help to constrain the gas accretion rate.
Peculiar galaxies without any companion are
quite frequent, about 50\% out of a sample of 1700 galaxies have
an HI asymmetric profile (Richter \& Sancisi 1994). 
The asymmetry is even more frequent for 
late-type galaxies, about 77\% (Matthews et al 1998).
The asymmetry is also observed in the
stellar disk in the optical light (Zaritsky \& Rix 1997).

Recently, the amplitude of lopsidedness has been 
quantified precisely in the NIR distribution 
(from 150 galaxies of the OSU sample) and compared to what is
expected from numerical simulations 
(Bournaud et al 2005).
The $m=1$ perturbations could be excited by a companion,
or during the formation of the galaxy, since they can
persist under the form of kinematic waves for quite a long
time, but still not sufficient to explain the observed
frequency now (Baldwin et al 1980).

It is difficult to explain all lopsidedness
through the tidal interaction from companions, since
most galaxies are isolated (Wilcots \& Prescott 2004),
and for those with visible companions, 
the amplitude of the $m=1$ perturbation
does not correlate
with the tidal index (proportional to the mass 
of the companion and inversely to the cube of its distance). 
In addition, the amplitude of the $m=1$ and $m=2$
perturbations are correlated, and also
the $m=1$ intensity correlates with type, which is
not predicted for galaxy interactions.

\begin{figure}[ht]
\centerline{ \begin{tabular}{cc}
\includegraphics[angle=0,width=6cm]{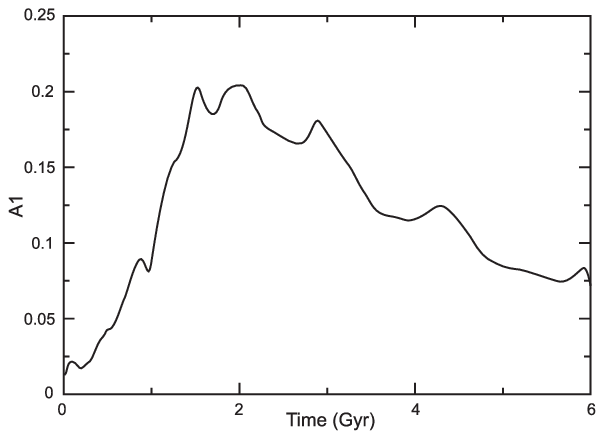}&
\includegraphics[angle=0,width=10cm]{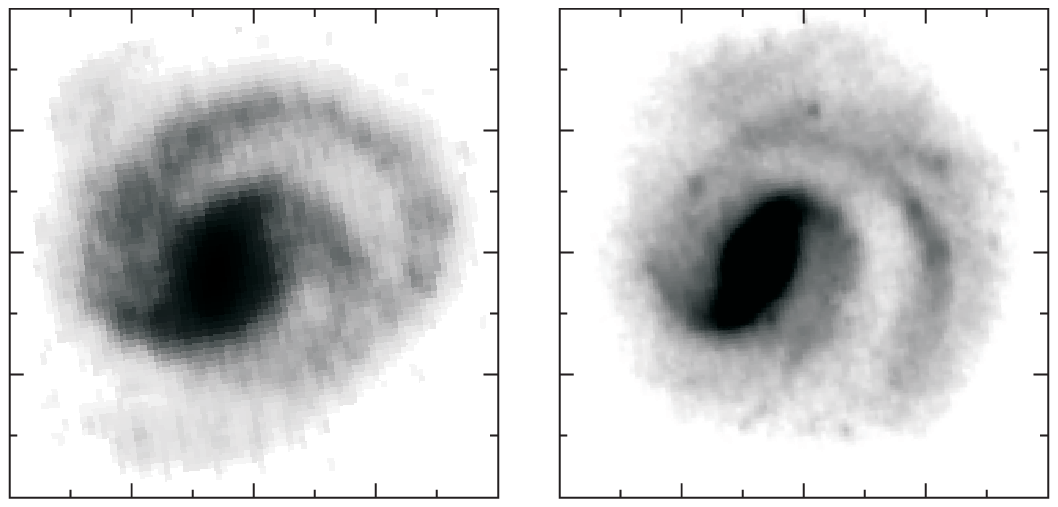}\\
\end{tabular}}
\caption{{\bf Left:} Evolution of $A_1$, the amplitude of the 
$m=1$ perturbation, in a simulation of asymmetrical gas accretion. A
 strong lopsidedness is triggered, even in the stellar component (through star formation),
in a galaxy that looks completely isolated. 
{\bf Middle:} Stellar image of  a simulation of asymmetrical gas accretion, 
with a rate of the order of  4~M$_{\odot}$~yr$^{-1}$, which produces
a lopsided galaxy quite similar to NGC~1637 
{\bf Right:}  NIR map of NGC~1637 from the OSUBGS data, after deprojection.
 A strong lopsidedness is present in this very isolated galaxy
(from Bournaud et al 2005).}
\label{accr}
\end{figure}

The cause of the $m=1$ could then be either a
minor merger (no companion is seen), or external gas accretion.
Both processes have been simulated, and the amplitude
of the $m=1$ perturbation then quantified, as a function of time
(Bournaud et al 2005). 	It is found that only asymmetric gas accretion
is able to reproduce the high frequency of the asymmetries observed,
since the result of a minor merger  becomes symmetric quite early (see Figure
\ref{accr}).
Only gas accretion (here with 4 M$_\odot$/yr)
can explain the observed frequency of $m=1$
and the long life-time of the perturbation in
NGC 1637.
A large number, at least two thirds, of strong $m=1$ galaxies
require external gas accretion.

%--------------------------------------------------------------------------
\section{ Environmental effects in clusters}

The relative importance of the hierarchical scenario
and secular evolution must depend
strongly on environment. In particular, in rich clusters, mergers have
considerably increased the fraction of spheroids and ellipticals, and
tidal interactions and ram pressure have stripped and heated the cold
gas around galaxies, so that external accretion will be reduced or suppressed.

Although massive galaxies in rich clusters are passively evolving
ellipticals, devoid of any star formation, this has not always been 
the case. Clusters have evolved in a recent past.
High resolution images with the HST, followed by spectroscopic
surveys, have shown that there exist signs of 
tidal interaction/mergers in z=0.4 clusters. There
is a much larger fraction of perturbed galaxies,
a larger fraction of late-type and starbursting
objects.  Rings of star formation are 
much more frequent than 2-arm spirals
(Oemler et al 1997). And it is well known that there is an excess
of blue galaxies as a function of z 
(Butcher, Oemler 1978, 1984).

Moreover $z \sim 0.5$ clusters
possess a large fraction of peculiar galaxies, likely post-starburst,
called E+A (or k+a), devoid of emission lines (and therefore with
no current star formation), but very strong Balmer absorption lines
(Dressler et al 1999, Poggianti et al 1999). This means that they have
a large fraction of A stars, implying that the galaxy was experiencing
a strong starburst that has just been abruptly interrupted.
Star formation was quenched, in 
these galaxies in majority disk-dominated. Their fraction is
about 20\%, much larger than in the field, and in the clusters at $z=0$
(Dressler et al 1999).

Star formation and morphological evolution 
appear decoupled in cluster galaxies (Couch et al 2001), while
star formation is still occuring actively now in groups (Balogh et al 2004).
The star formation rate is strongly dependent
on the local projected density, star formation is quenched as soon as
the density is above $\Sigma$ = 1 galaxy/Mpc$^2$, independent of the size
of the structure. It is likely that the bulk of the stars now in massive
galaxies have been formed in spiral galaxies in groups, before merging
into ellipticals. 

%--------------------------------------------------------------------------
\section{Conclusion}

Three main processes have been invoked for the 
 mass assembly of galaxies: monolithic collapse (MC),
hierarchical merging (HS) and
external gas accretion associated with secular evolution (SE).
  The MC scenario appears limited at high redshift,  involves
only sub-components of galaxies and has a limited role, while the
two others (HS and SE) compete with comparable weights to galaxy
formation.

Observations constrain the relative importance of these processes.
The star formation history across the Hubble time, either through
the fossil record of present stellar populations, or direct observatons
of galaxies at various redshifts, reveals that massive galaxies have
now stopped their star forming activity, while intermediate-mass spirals
are still continuing with an almost constant rate, and dwarf galaxies
experience starbursts. The observation of massive galaxies at high redshift
is compatible with the predictions of the $\Lambda$CDM model of galaxy
formation.
The observation of the dynamical state of galaxies 
(bars, spirals, warps, polar rings, $m=1$ asymmetries) also constrain
the importance of external gas accretion, and secular evolution.

The different processes play a different role, according to the environment.
In the field, gas accretion is dominant, to reform bars and spirals, to
explain warps, polar rings or lopsidedness.
In rich environments, the evolution proceeds at a much faster pace, 
hierarchical merging had much more importance, and 
secular evolution of galaxies is halted at z $\sim$ 1, since galaxies
are stripped from their gas, and from their cold filamentary structure,
acting as a gas reservoir. 

%--------------------------------------------------------------------------
\section*{Acknowledgments}
Many thanks to the organisers, and in particular
the chairmen D. ELbaz and H. Aussel, for inviting me to
such a nice and fruitful conference.

%--------------------------------------------------------------------------

\end{document}